\documentclass{an}
\usepackage{times,epsf,graphicx}
\usepackage{natbib}
\bibpunct{(}{)}{;}{a}{}{,}

\def\sol{\ensuremath{_\odot}}

\newcommand{\teff}{\ensuremath{T_{\rm{eff}}}}
\newcommand{\logg}{\ensuremath{\log g}}

\begin{document}

\Pagespan{1029}{}
\Yearpublication{2010}
\Yearsubmission{2010}
\Month{11}%
\Volume{331}%
\Issue{9/10}%
\DOI{10.1003/asna.201011450}%

\title{Observational Asteroseismology of Hot Subdwarf Stars}

\author{R.~H.~{\O}stensen}

\titlerunning{Observational Asteroseismology of Hot Subdwarf Stars}
\authorrunning{R.~H.~{\O}stensen}
\institute{
  Instituut voor Sterrenkunde, K.U.Leuven,
  Celestijnenlaan 200D, B-3001 Leuven, Belgium
}

\received{2010 May 6}
\accepted{2010 Sep 13}
\publonline{2010 Nov 2}


\abstract{Hot subdwarf stars are particularly challenging for asteroseismology
due to their rapid pulsation periods, intrinsic faintness and
relative rarity both in the field and in clusters.
These features have ensured that the preferred method
of observation up to now has been white-light photometry, and all
asteroseismological solutions to date have been made by model fitting
of the frequency spectrum. Several attempts have been made to
perform asteroseismology using time-resolved spectroscopy on the brightest
of these stars, but with modest results. A few attempts at simultaneous
multi-color photometry have also been made to identify modes
with the amplitude ratio method.
We will review the most recent observational results and progress in
improving the observational methods for ground-based asteroseismology
of these compact pulsators.  }

\maketitle

\section{Introduction}

Hot subdwarf stars are mostly Extreme Horizontal Branch (EHB) stars,
{\em i.e.} core helium burning stars with an
envelope too thin to sustain hydrogen-shell burning. Not all sdB stars are
EHB stars, since evolutionary tracks of post-RGB stars that
fail to ignite helium also cross the subdwarf region on their
way to the helium-core white dwarf (WD) stage.
While the EHB term implies core He burning, the sdB/sdO terms
are used to describe the spectroscopic appearance, and do not presume a
particular evolutionary stage.

The canonical picture of the EHB stars was established
by \citet{heber86}, in which the EHB stars are He core
burning stars with masses close to the core He flash mass
of $\sim$0.47\,M\sol, and hydrogen envelopes too
thin to sustain hydrogen burning (less than $\sim$1\% by mass).
It is understood that they are post-RGB stars that have ignited
He in a core flash just before or after the envelope
was removed by any of several possible mechanisms.
The lifetime of EHB stars from the zero-age EHB (ZAEHB) to the
terminal age EHB (TAEHB), when core He runs out, takes between
100 and 150\,Myrs, and the post-EHB evolution will take them through the
sdO domain directly to the WD cooling curve without ever passing through
a second giant stage. The time they spend shell He burning before
leaving the sdO domain can be up to 20\,Myrs \citep{dorman93}.

Although the future evolution of EHB stars after core He exhaustion has
always been presumed quite simple, the paths that
lead to the EHB are still somewhat mysterious.
New hope that the evolutionary paths leading to the formation of
EHB stars can be resolved has been kindled by the discovery that many
of them pulsate, which has opened up the possibility of
probing their interiors using asteroseismological methods.
For an introduction to the interior structure of the EHB
stars, and the pulsation driving mechanism we defer the reader to
the accompanying review paper by Kawaler (this issue).
Here we will focus on the observed properties of
EHB stars, and on the particular challenges for observational
asteroseismology.

Hot subdwarf stars are often found as blue stars in surveys covering
the galactic caps.
The PG survey \citep{pgcat} covered more than ten thousand square
degrees at high galactic latitudes, and found 1874 UV-excess objects, of
which more than 1000 were identified as hot subdwarfs,
so these stars dominate the
population of faint blue stars down to the PG survey limit ($B$\,=\,16.5).
Together with the large sample of subdwarfs detected in the HS survey 
and analysed by \citet{edelmann03},
these have provided a rich source of hot subdwarfs
for observers to follow up, and discoveries of new variables from these
surveys are still made almost every year \citep[see][and references therein]{sdbnot}.
The recent SDSS \citep{SDSS}
also contains spectra of more than 1000 hot subdwarfs, but is
dominated by the increasing fraction of faint galactic WD stars, which
takes over as the dominant population
around about $B$\,=\,18 mag, at a distance which hot subdwarfs belong to the
dilute halo population.
The Subdwarf Database \citep{ostensen06} lists $\sim$2500
hot subdwarfs, with extensive references to the available literature.

Several surveys have attempted to tackle the question of the binary
frequency of EHB stars, but the matter is complicated by the many different
types of systems in which these stars are found. Hot subdwarfs with FGK companions
are easily detected from their double-lined spectra or from IR excess.
But such stars with WD or M-dwarf companions show no such features.
When the orbital periods are sufficiently short, these systems can
easily be revealed from their radial velocity (RV) variations.
Using the RV method, \citet{maxted01}
targeted 36 sdB stars and found 21 binaries, all with periods less than
30 days. This gave a fraction of short period binaries of 60\,$\pm$\,8\,\%.
Other surveys have found smaller fractions, but they have not constrained
the sample to focus strictly on the EHB.
From high-resolution VLT spectra of 76 sdB stars
\citet{lisker05} found that 24 
showed the signature of an FGK companion, none of which showed any
detectable RV variability. \citet{napiwotzki05} reported that of 46 sdB stars in
the same sample, 18 (39\%) were RV variable. Clearly, the binary fraction
in EHB stars is substantially higher than for normal stars,
but the exact number is hard to pin down since sample selection and
survey sensitivity effects easily skew the numbers.

\begin{figure}[t]
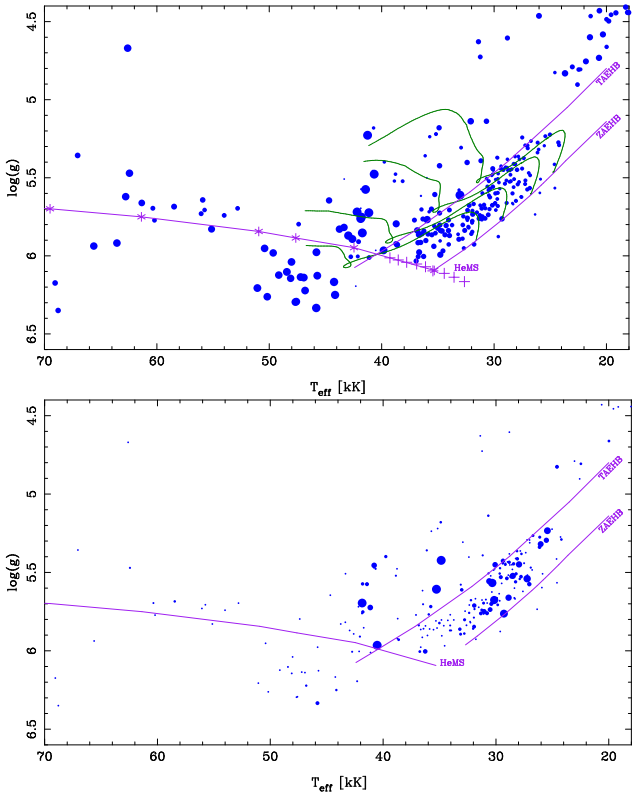

\includegraphics[angle=0,width=\hsize]{BG_TGHe+EHB.eps}
\includegraphics[angle=0,width=\hsize]{BG_TGRV.eps}
\caption{
The EHB in the \teff--\logg\ plane as observed by the Bok-Green survey
(Green et al.~2008).  The symbols mark observed stars with the
size indicating the He abundance.
The theoretical zero age HeMS is shown for a wide range of masses.
Models from Paczy\'nski (1971) with masses of 0.5, 0.7, 0.85, 1, 1.5
and 2\,M\sol\ are marked with $\ast$ symbols (starting from low \teff).
More recent models from \citet{kawaler05} are shown
for M$_\ast$ = 0.41, 0.43, ... 0.57 M\sol\ and
marked with + symbols. The ZAEHB and TAEHB
for 0.47\,M\sol\ core models are also drawn.
For the latter, four evolutionary tracks with different envelope
thicknesses $\log M_e/M_\ast$\,=\,-3.5,-3,-2.5,-2 are drawn
(starting from high \logg ). 
Lower panel: As above, but with the symbol size indicating the dispersion
in RV. The EHB stars with the highest velocity
variations appear to be concentrated at lower gravities on the EHB.
sdB+FGK stars are not included here,
due to difficulties in reliably disentangling composite spectra.
Figures from \citet{ostensen09}.
}
\label{fig:TG_RV}
\end{figure}

Most recently, \citet{green08} presented a uniform high
signal-to-noise low-resolution survey of a substantial sample including most
known hot subdwarf stars brighter than $V$\,=\,14.2, using the university
of Arizona 2.3\,m Bok telescope (hereafter referred to as the
Bok-Green or BG survey).  From this large sample the clearest
picture of the EHB to date emerges (Fig.\,1).
Most stars in the diagram are clearly well bound by EHB models
for a narrow mass distribution.
Most of the remaining stars are consistent with post-EHB models, but could
also fit core helium burning objects with higher than canonical masses.
The most helium rich objects, however, appear to form their
own sequence, which cannot be explained by canonical EHB models.
Although the details of this survey are still under analysis, several
new features have been noted. The sequence of He-rich objects around
40\,kK is not compatible with current evolutionary scenarios,
since post-EHB and post-RGB objects pass too rapidly through this region
of the \teff--\logg\ plane to produce the observed clustering, but
the {\em late hot flasher} scenario \citep{sweigart97}
holds some promise.

\begin{figure}[t]
\includegraphics[angle=0,width=\hsize]{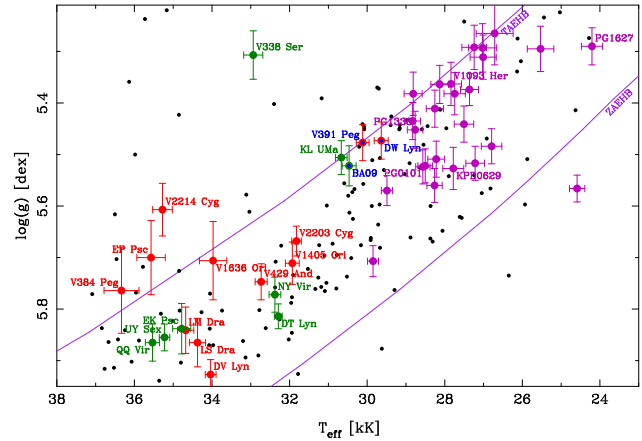}
\caption{
Section of the \teff--\logg\ plane where the EHB stars are located.
Pulsators with temperatures and gravities in the BG survey are marked
with big symbols and error bars. Small symbols without error bars are
stars not observed to pulsate.
In the online version the colors indicate
short period pulsators with ({\em green}) and without ({\em red}\/)
published asteroseismic solution, long period pulsators ({\em magenta})
and hybrid pulsators ({\em blue core}).
}
\label{fig:EHB}
\end{figure}

\begin{figure*}[t]
\centerline{
\includegraphics[angle=-90,width=8.5cm]{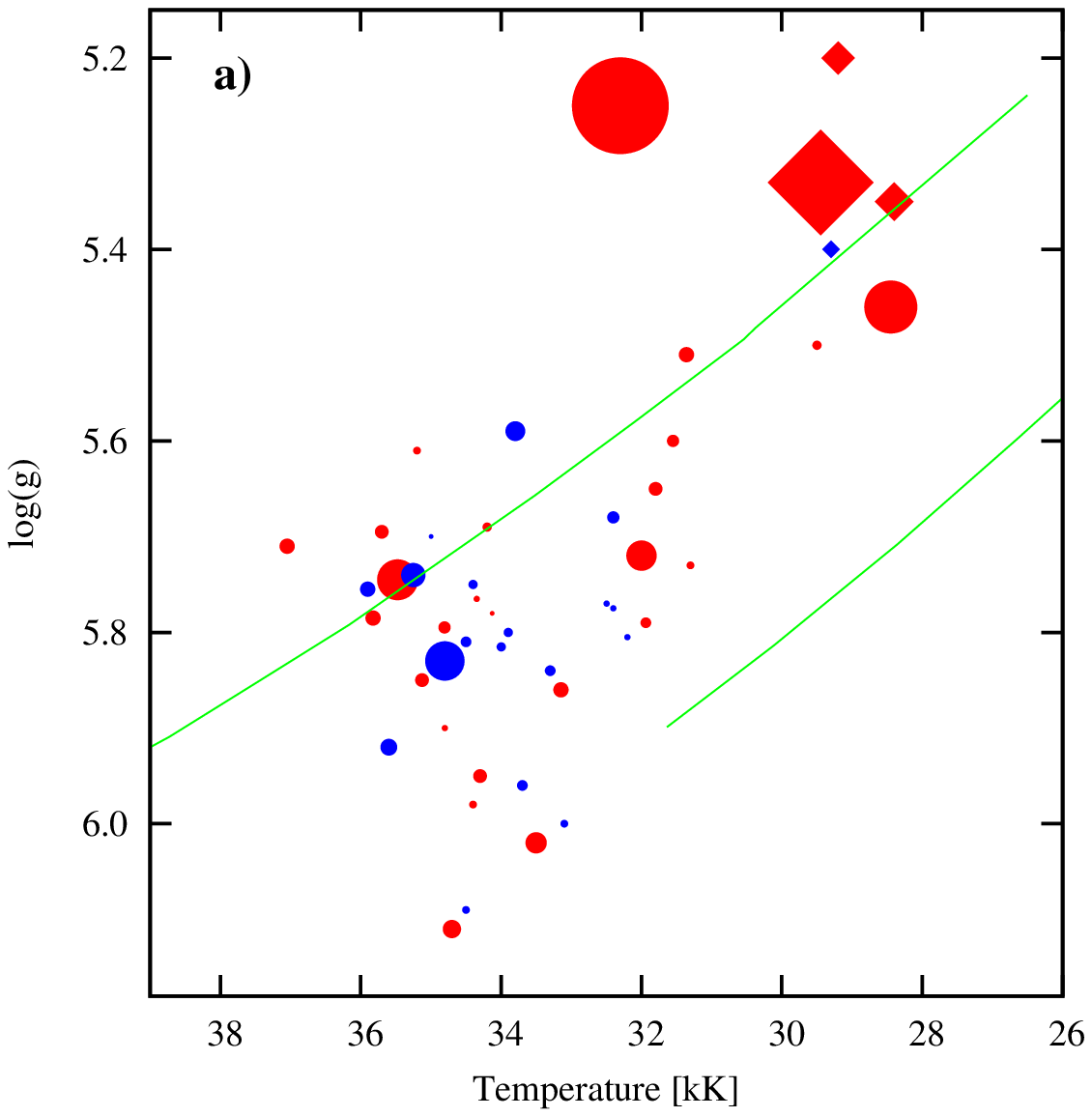}
\includegraphics[angle=-90,width=8.5cm]{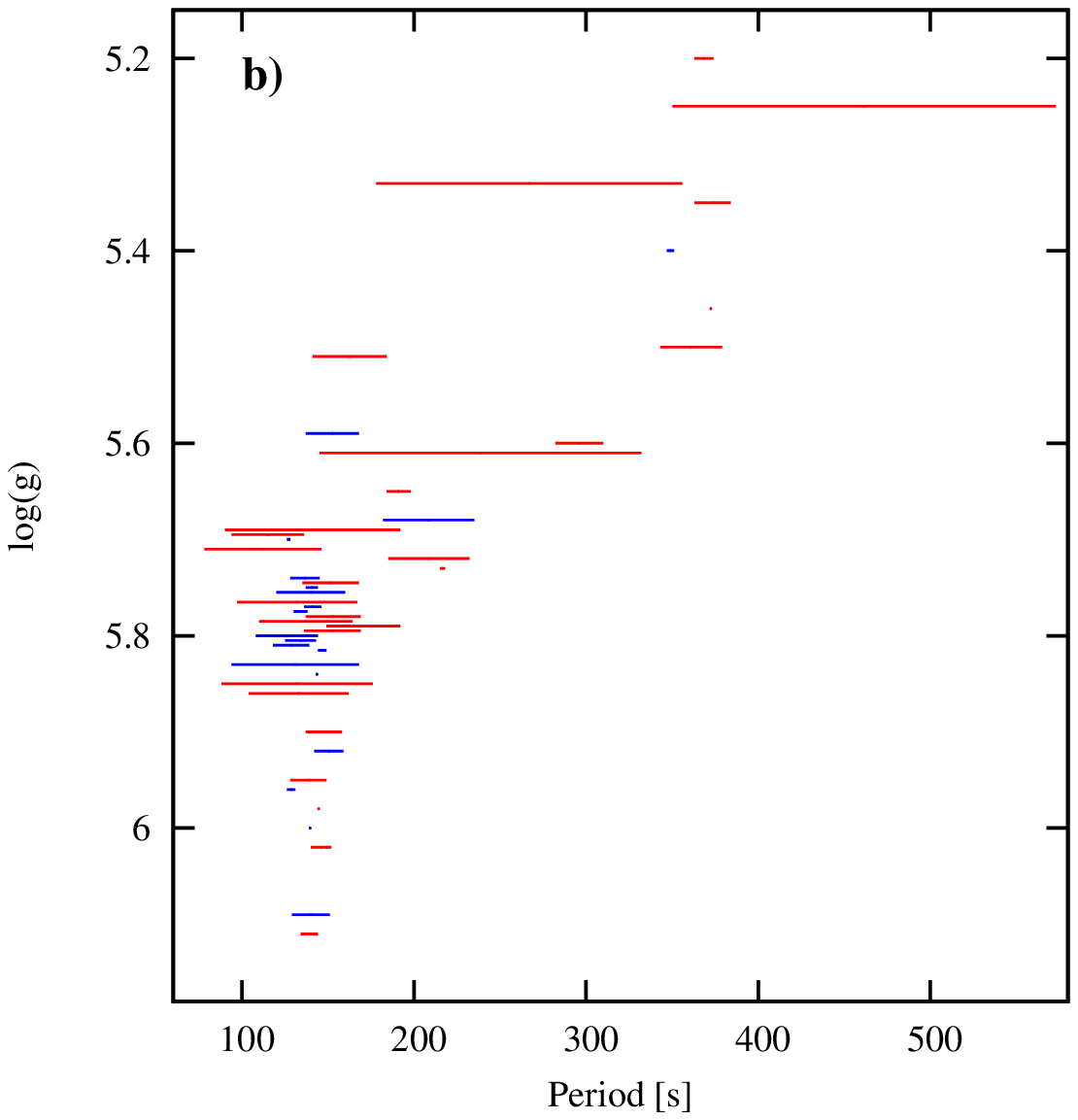}
}
\caption{
The properties of the 49 known sdBV stars are summarised in
\citet{sdbnot}. These figures appear in that paper, and show:
a) \teff--\logg\ diagram for the 49 sdBVs.
Circles indicate V361\,Hya type pulsators and diamonds indicate
hybrid DW\,Lyn type pulsators, and the size of the symbols is
relative to the pulsation amplitude. The 19 sdBVs from the NOT
survey are shown in blue, and pulsators from other surveys in red.
The green lines indicates the EHB, as in Fig.~\ref{fig:TG_RV}.
b) $P$--\logg\ diagram for the same pulsators.
The bars indicate the range of detected pulsation periods in any particular
star, excluding harmonics and $g$-modes.
}
\label{fig:tgp}
\end{figure*}

The binary interaction scenarios that permit a star on the
RGB to be stripped of its envelope to leave a helium burning core that
will settle on the EHB are quite well understood \citep{han02,han03}.
Depending only on the mass ratio of the system, the orbit will either contract
or expand. 
If the expanding mass donor is more massive than the accretor, the orbit will
shrink catastrophically and the system enters a common envelope (CE) phase.
As the orbit shrinks further due to friction, orbital energy is deposited
in the envelope, spinning it up and eventually the envelope will be ejected.
If the companion is more massive than the RGB donor, the orbit expands and no
CE is formed. In this stable Roche lobe overflow (RLOF) scenario
the orbital period can end up as long as 2000 days.
The formation of single sdB stars is more problematic. A possible formation
path is the merger between two helium white dwarfs first proposed
by \citet{webbink84}. This can easily produce
many of the He-rich subdwarfs, but it is hard to understand how sufficient
hydrogen can survive for them to end up with significant envelopes, as
most EHB stars have. The enhanced stellar wind models proposed by
\citet{dcruz96} are a possibility, but it is hard to explain why RGB 
stars should display such a large variability in wind mass loss rates
that is required to explain the observed distributions.
For a discussion of the more speculative scenarios, see \citet{ostensen09}.

\section{Asteroseismology}

Rapid pulsations in sdBs were first reported by
\citet{kilkenny97}, after their discovery of the prototype,
V361\,Hya (EC\,14026-2647), and several similar objects.
These pulsators span the hot end of the EHB strip
and pulsate in $p$-modes of low 
$\ell$ orders with photometric amplitudes up to 6\%.
The pulsation periods range between 100 and 400\,s, and 49 such
stars are known in the literature to date \citep{sdbnot}.
One of these, V338\,Ser (PG\,1605+072),
has periods reaching almost 10 minutes, but stands out as it
sits well above the EHB (Fig.~\ref{fig:tgp} and \ref{fig:EHB}),
possibly because it is in a post-EHB stage of evolution.

Long period pulsations became known when
\citet{green03} reported pulsations in V1093\,Her
(PG\,1716+426), with periods between one half and two hours.
They also found that as many as 75\% of sdB stars cooler
than $\sim$30\,kK display some level of pulsations at these periods,
but due to atmospheric effects at similar time-scales
and strong multi-periodicities, they are hard to reliably
characterise.
These stars span the EHB from the coolest sdBs up to the domain
of the V361\,Hya stars (Fig.~\ref{fig:EHB}).
The pulsations were identified with
high radial order $g$-modes by \citet{green03}, and their
amplitudes are very low, typically 0.2\,\%.

When \citet{schuh06} found that a known V361\,Hya star, DW\,Lyn
(HS\,0702+6043), was
displaying the $g$-modes of a V1093\,Her star simultaneously with
short period $p$-modes, the existence of hybrid pulsations
was established. Such stars are now often referred to as DW\,Lyn stars.
A final class of pulsations in hot subdwarf stars
was discovered by Woudt et al.~(2006) in the hot sdO/F-star binary
J17006+0748.
% Having a close F-star companion makes it very likely that
% the sdO star was formed via the stable RLOF channel, and is now
% in a post-EHB shell burning stage.  Its effective
% temperature is around 65\,kK, the gravity appears to be around
% \logg\,=\,5.5 and the atmospheric He abundance about twice of
% solar from both the SALT spectrum presented by \citet{woudt06} and the
% SDSS survey spectrum.
% The frequency range detected in this star is from 1 to 2 minutes,
% which is even shorter than the V361\,Hya stars, and the
% amplitude of the main mode is about 4\%, comparable to the highest
% amplitude V361\,Hya stars. J17006+0748 is very faint at $g$\,=\,17.4,
This object is very faint at $g$\,=\,17.4,
and no other pulsating sdO star has yet been reported. 

For a discussion on the driving mechanism and interior structures
of hot subdwarf pulsators, we refer the reader to the article by
Kawaler (this issue). In this review we will focus on the methods
that have been used to identify pulsation modes in the
sdBVs, which will be described in the following sections.

\begin{figure*}[t]
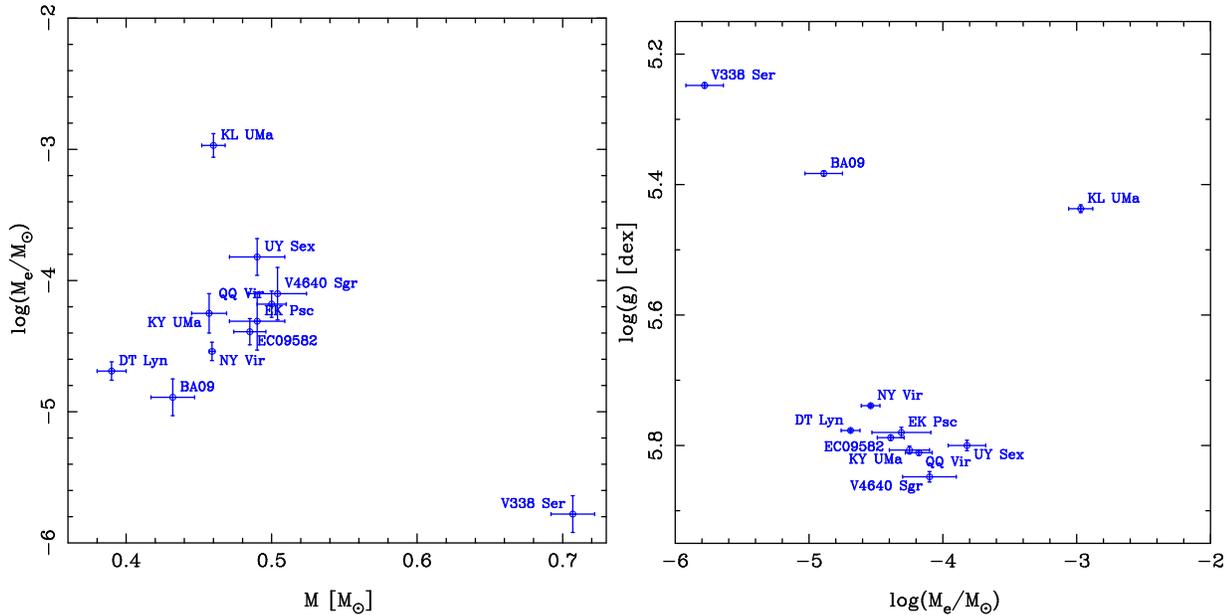

\centerline{
\includegraphics[angle=0,width=8.0cm]{seismo_MMe.eps}
\includegraphics[angle=0,width=8.1cm]{seismo_MeG.eps}
}
\caption{
Published asteroseismic solutions for ten V361\,Hya stars.
The left hand panel shows total mass plotted versus envelope mass,
and the right hand panel shows envelope mass versus surface gravity.
Note that only the optimal solution is shown even if the
papers discuss several possible ones.
}
\label{fig:seismo}
\end{figure*}

\subsection{Period matching}

The exceptional amplitude of the dominant period in Balloon 090100001 has
hinted towards a radial nature, and several mode identification methods
have now confirmed that suspicion.
\citet{vanGrootel08} have successfully applied the forward
method to Balloon 090100001,
demonstrating some peculiarities in the model predictions.
Their optimal solution for the main mode, when using no constraints,
is $\ell$\,=\,2, which is not reconcilable with the spectroscopic data.
However, by imposing mode constraints from multicolor photometry they do
find asteroseismic solutions that agree with all observational data.
Curiously, the physical parameters for the constrained and unconstrained
fits are almost identical, even if the mode identification changes for
half the modes considered.  This peculiarity arises from the high
mode density and the way the modes are distributed in period space.

With a recent update of the forward modelling code, \citet{charpinet08}
have produced a very convincing model for the eclipsing binary
system NY\,Vir (PG\,1336-018).
This star has been particularly challenging since it is rapidly rotating,
due to being in a tidally locked orbit with the close M-dwarf companion.
The rotational splitting of modes with different
$m$ produces a particularly rich pulsation spectrum.
\citet{charpinet08} use asteroseismology to discriminate between
three solutions from the binary orbit published by
\citet{vuckovic07}, and finds that the intermediate model
with a mass of 0.47\,M\sol\ is clearly favored. This solution is
also the only one consistent with the \logg\ from the BG survey
(Fig.~\ref{fig:EHB}).

To date, eleven asteroseismic solutions computed with the forward method have
been published.  They were summarised in \citet{randall07} for the
first seven, and in Fig.~\ref{fig:seismo}
the new solutions by \citet{vanGrootel08}, \citet{vanSpaandonk08},
\citet{charpinet08} and \citet{randall09b} have been included.
A feature of the asteroseismic modelling is that \teff\ is rather
poorly constrained, and a better value can usually be
provided from spectroscopy. The surface gravity, total mass, and envelope
mass fraction all have very small associated errors in the
asteroseismic solutions, so we plot only these in Fig.~\ref{fig:seismo}.
The distribution of masses is not as concentrated around 0.47\,M\sol\ as
most canonical evolutionary models have presumed,
but all points are well within the permitted ranges for synthetic
populations considered by \citet{han02,han03}.
Except for two outliers, all the stars
appear to form a trend with envelope mass, $M_{\mathrm e}$,
increasing with total mass, $M$.
Although this feature has not been accounted for by evolutionary
calculations, it could occur as a natural consequence of a 
higher core mass requiring more energy to remove the envelope.
More disturbing is the lack of any clear trend in envelope mass
versus surface gravity, as is clearly demanded for canonical EHB models.
The scatter in the high gravity objects is easily explained by their spread
in mass, and KL\,UMa fits well with the expected $M_e$/\logg\ trend. But
the unusually low envelope masses for BA09 and V338\,Ser are hard to
explain, and may indicate that the adopted models are too simplified
to represent the seismic properties for these cases.

There are a few concerns with the forward matching method as used on the
V361\,Hya stars up to now. First of all, the models only match the observed
frequencies to a precision of a few percent. The frequencies can be determined
to within a $\mu$Hz in a few weeks of observations (which is one fiftieth of
a percent for typical short-period sdB pulsations), and a precision of
0.02\,$\mu$Hz was reported for V391\,Peg by \citet{silvotti02b}.
Much more work on theoretical models are required to bring them
closer to this precision level. 
This requires non-adiabatic pulsation models to be connected to fully evolutionary
models for sdB stars. Such evolutionary models must include gravitational settling
and radiative levitation. Progress on the first point was reported recently
by \citet{hu09}, but the second point still remains to be solved.
Another issue with the current model calculations is that they do not
address the pulsation amplitudes of the modes.

Using photometric mode-ID has a distinct advantage on faint stars,
but requires a rich pulsation spectrum.
Multi-site campaigns can provide high frequency resolution and detect
low amplitude modes. However, often, no unique solution for a particular
mode is found, because modes with different $\ell$'s can be found within
the precision of the model fit.

V1093\,Her stars are much harder to do with this method, since the longer
periods requires correspondingly long time-bases in order to resolve the
pulsation spectra.
Only one long baseline study has been attempted; PG1627+017 \citep{randall06b}.
From 300\,h of photometry they found 23 periods with amplitudes
between 0.4 and 5 mma.
However, as gravity modes probe the interior of these stars, reliable mode
identification from period matching
can only come from models that include detailed interior structure.

\subsection{Spectroscopic mode ID}

\citet{telting08} presented the first study of line-profile variations in
these stars based on high-resolution spectroscopy, again using Balloon 090100001
as the preferred target due to its strong dominant mode.
While the line-profile method is well established for various main sequence
pulsators, its application to the faint sdB stars requires substantial
investments in terms of telescope time, which has hampered its use.
With the preliminary results on Balloon 090100001,
\citet{telting08} demonstrated
that the $\ell$ of the main mode must be either zero or one.

It is possible to make a direct mode-ID of a single pulsation period from
phase diagrams of sharp spectral lines, but that method requires both high
S/N and that the star has a significant rotation. Except for a few sdB
stars that are in short period binaries and therefore have tidally locked
rotation, all sdB stars studied so far shows extremely low rotation
velocities ($v\sin(i)$\,$\le$\,5\,km/s), which explains why this method
have not been successful yet on these pulsators.

The main problem with using spectroscopic mode identification techniques
on the hot subdwarf pulsators is the extraordinary effort in terms of
telescope size and time required in order to 
get the required S/N. So far it has only been feasible on a few pulsators
with a high amplitude main mode, and even then phase folding of extensive
time series has been required.
In multi-mode pulsators with many peaks of comparable amplitude, interference
between the modes produce broadening in the phase folded line profiles
which further hamper their interpretation.
But progress is still being made both in the observational methods and in
the methods used to interpret the observations.

\subsection{Amplitude ratios from multi-color photometry}\label{multicol}

The amplitude ratio method allows unique identification of $\ell$ for
all modes in a pulsation spectrum.
This is done by comparing the ratio of pulsation
amplitudes as observed in different photometric bands, with amplitudes
computed from models that include limb darkening effects.
The method requires very high accuracy on the amplitudes
in order to distinguish between $\ell$\,=\,0,\,1,\,2 modes \citep{ramachandran04},
but can easily distinguish between the low $\ell$ modes and $\ell$\,=\,3,\,4,\,5.
To achieve the required precision, simultaneous observations in several bands
are required. The clever design of {\sc ultracam} \citep{dhillon01}, in which
dichroics are used to split the light into three beams that permits truly
simultaneous observations in three passbands, has made such observations
possible.  {\sc ultracam} observations on V361\,Hya stars were first obtained
by \citet{jeffery04}, where they demonstrated that modes of $\ell$\,=\,3 and 
4 must both be present in their stars in addition to $\ell$\,$\le$\,2 modes, but
they did not have sufficient data to distinguish between the low $\ell$ modes.
More recently \citet{vuckovic10} presented six nights of {\sc ultracam} photometry
of EO\,Ceti (PB\,8783) which does have the required accuracy for mode-ID. However,
in this case the result is hampered by the fact that EO\,Ceti has a strong F-star
companion, which gives a light-contribution that is more severe in the red part
of the spectrum, and therefore skews the amplitude ratios. But they did manage
to prove that the photometric amplitude ratios remain constant even when the
amplitudes themselves appears to vary in time, which is encouraging as such
amplitude variability have been observed in just about all V361\,Hya stars
\citep{kilkenny10a}.

\subsection{Combination method: multi-color + RV}\label{rvcomb}

Amplitudes from multi-color photometry can be used in combination with radial
velocity amplitudes from spectroscopy in order to obtain more reliable discrimination
between low $\ell$ order modes. The method was first introduced by
\citet{jagoda03}, and applied to Balloon 090100001 by \citet{baran08}, where they
obtained a clear preference for $\ell$\,=\,0. 
More recently, \citet{baran10b} applied the method to QQ\,Vir, and again found a
preference for $\ell$\,=\,0 for the dominant main mode, but in this case the
discrimination is not clear enough to exclude $\ell$\,=\,1.
This method has a clear advantage in that it does not require such an extreme
S/N level as the multi-color method alone. The catch is that the spectroscopy
must be obtained close to the same time as the photometry in order to ensure
that the pulsation amplitude has not changed between the times of the photometric
and spectroscopic observations.

\begin{figure*}[t]
\includegraphics[angle=0,width=10cm]{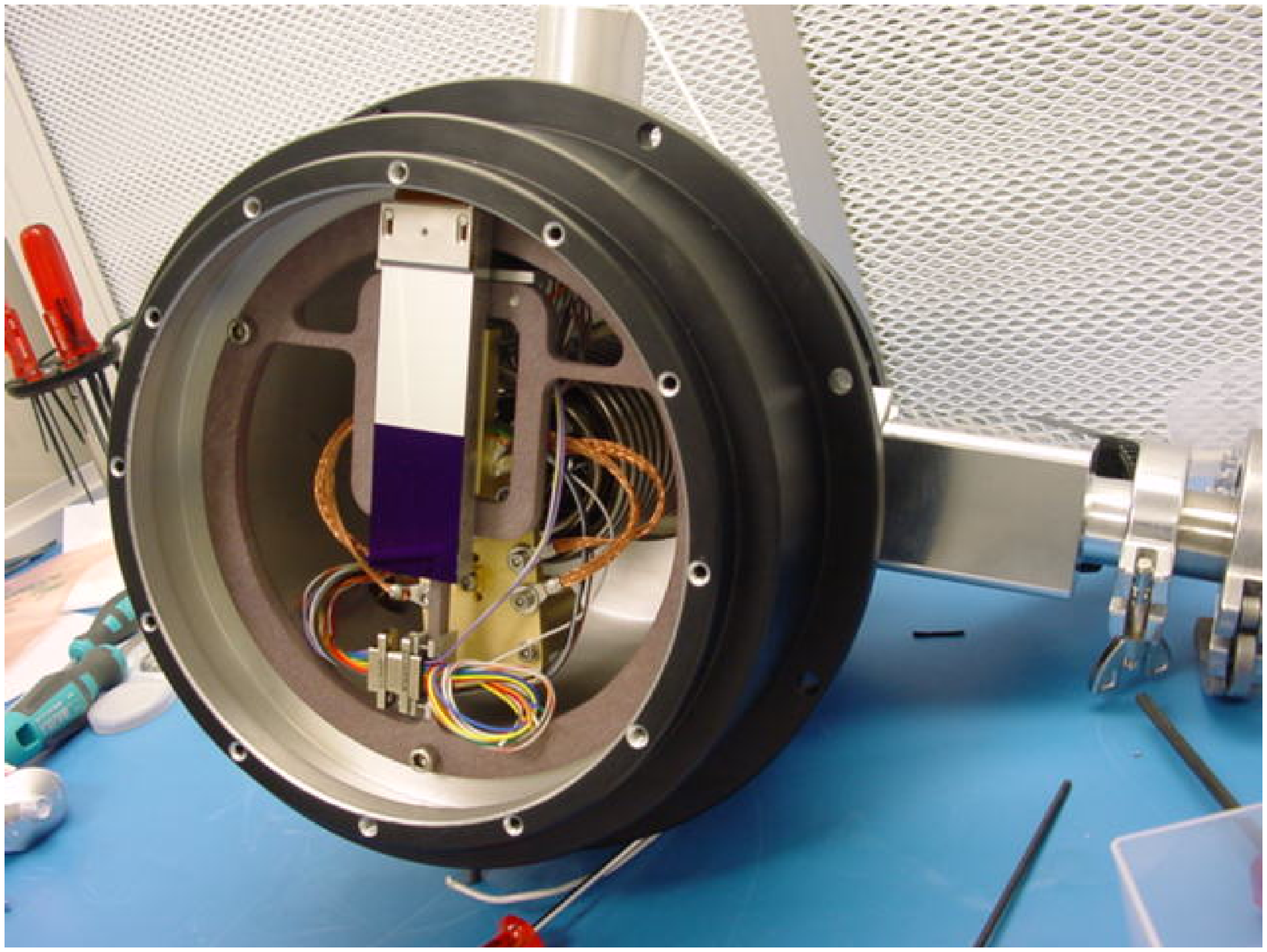}
\includegraphics[angle=0,width=7cm]{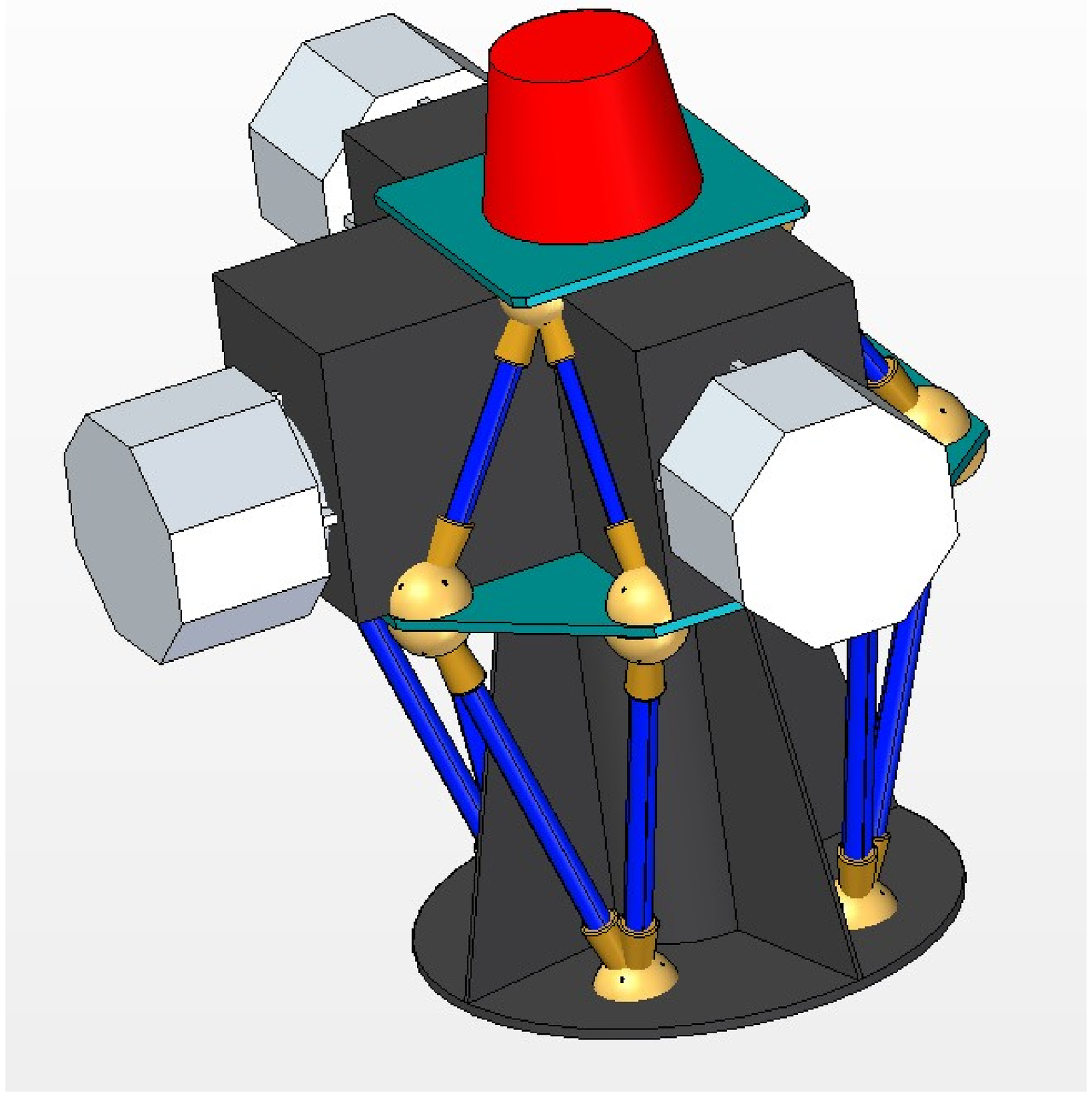}
\caption{
{\sc Left}: The upgraded {\sc merope ii} camera during the mounting of the
large format frame-transfer CCD at the {\em Mercator Telescope}.
The 13\,cm long CCD package requires a specially designed asymmetric
cryostat design due to the extensive image storage area (silver coated),
clearly visible above the anti-reflection coated imaging area.
{\sc Right}: The design for the three-channel {\sc maia} camera will apply
three such CCDs in separate cameras kinematically mounted in the Nasmyth
focus of {\em Mercator}.
}
\label{fig:maia}
\end{figure*}

\section{Progress in observations}

Multi-site campaigns have been organised to observe sdB
pulsators since the time of the first discoveries \citep{kilkenny99},
in order to reduce one-day aliases in the Fourier spectrum.
But combining data from different observatories
using different filters is a method that is riddled with problems, when
high precision is required not just on the frequencies but also on the
amplitudes. 
Since the amplitudes of the pulsations are stronger in the blue,
individual modes may appear to have different amplitudes for
observers with different passbands. This is not trivial to correct for
since the color dependence of the pulsation amplitudes also depends
on the order, $\ell$. The only way to obtain continuous time series photometry
unaffected by the diurnal cycle and atmospheric extinction effects is
to observe from space.

\subsection{The future is in space}

Data from the {\em Kepler Mission} are already
revolutionising the prospects of white light asteroseismology
on the hot subdwarfs. The unprecedented precision and exceptionally
high frequency resolution obtainable by space based photometry will
certainly have a major impact on our understanding of the hot subdwarfs.
While asteroseismology of the V361\,Hya stars have proven successful in
many cases, only with the new range of precision and resolution
obtainable from space can we expect to achieve successful asteroseismology
of the of V1093\,Her pulsators for the first time. % \citep{randall06}.

The only problem with {\em Kepler} is its limited field of view, which will
provide a restricted number of targets to work with.
The field is expected to contain about four V361\,Hya stars and perhaps eight
V1093\,Her stars \citep{silvotti04}.
% But this assumes that the 142 compact pulsator
% candidate stars that have been scheduled for short cadence observations with
% {\em Kepler}, contains all the stars in the field that we have expected,
% from the space density of these stars.
A spectroscopic survey of the candidate subdwarf pulsators have already been
completed, and will be presented together with the first {\em Kepler} results
in \citet{kepI}. While the number of pulsators expected is quite small,
at least for the V1093\,Her stars the sample should be sufficiently large
that we are certain to enter a new era of asteroseismology that will
fundamentally change our understanding of the subdwarf B stars.
But for all the known V361\,Hya stars in the literature, our observation tools
are rooted in ground-based facilities for a long time yet.

\subsection{Better multi-color photometry}

{\sc ultracam} combines two key design features that makes it uniquely
suited for mode identification of the rapidly pulsating hot subdwarf stars.
The first, as mentioned earlier, is the splitting of the incoming light beam
into separate channels, which allows truly simultaneous
multi-band photometry. The second is the use of frame-transfer CCDs, which
eliminates the dead time associated with the readout period, which is typically
between 20 and 80 seconds for reading out the full frame on a regular CCD chip,
comparable to the pulsation period of the V361\,Hya stars.
The frame-transfer concept divides a CCD chip into an integrating half and an
image storage area, so that one frame can integrate while the previous frame
is being read out. {\sc ultracam} is the only instrument that combines both
these features. The only catch with frame-transfer CCDs, and the reason why
they are avoided by most new CCD cameras is that they are only commercially
available in relatively small format. The three {\sc ultracam} CCD chips have
only 1024$\times$1024 pixels, while the standard for astronomical CCDs have for
a long time been at least 2048$\times$2048 pixels. This means
that {\sc ultracam} has a smaller field of view than ideal (assuming a typical
pixel size of 0.2--0.3 arcsec/pixel), and this can
impact seriously on the precision of photometric amplitudes. For instance,
\citet{vuckovic07} reported that they were unable to locate a reference
star with a significant brightness in the $u$-band in the vicinity of
NY\,Vir when observing with {\sc ultracam} on the VLT, and were therefore
unable to calibrate the time-series properly. While the light-curve could
be recovered by using a scaled version of the $g$-band light-curve for
the differential photometry, this method may adversely impact the precision
required for the amplitude ratio method, as the $u$-band is the one that
provides the largest discrimination between modes.
Another problem with {\sc ultracam} is that the camera is only available
for a very limited number of nights each year, as it is a traveling visitor
instrument.

Motivated by the success of the {\sc ultracam} design the instrumentation
group in Leuven have made an
improved design for a three-channel camera, to be permanently mounted on
the {\em Mercator Telescope} at Observatorio del Roque de los Muchachos
on La Palma. This telescope is owned and operated by K.U.~Leuven, and
through our ongoing instrumentation program we have in 2009
commissioned a high-resolution multi-fibre spectrograph, {\sc hermes}.

The development of new cameras ideally suited for high-speed CCD photometry
was made possible by the unfortunate termination of the
{\sc esa/eddington} asteroseismology space mission.
At the time the decision to cancel the mission was made, E2V had already
designed and developed new frame-transfer CCDs for {\sc eddington}.
After an application to {\sc esa}, we obtained four of the prototype CCDs
on a permanent loan.  These CCDs are large format frame-transfer
devices with an image format of 2048$\times$3074 pixels, permitting a six
times larger effective area than currently available frame-transfer devices.
The first of these devices was installed in a regular single-channel
CCD camera replacing the aging {\sc merope} camera in the Cassegrain
focus of {\em  Mercator} (Fig.~\ref{fig:maia}).
The remaining three will be installed in
the {\em Mercator Advanced Imager for Asteroseismology} {\sc maia},
a three channel CCD camera of comparable design to {\sc ultracam},
which is currently being built in Leuven, and scheduled for commissioning
at {\em Mercator} next year \citep{MAIA-SPIE}.

\section{Conclusions}

Astrophysics of EHB stars is a rapidly advancing field with
exceptional challenges, due to their complex formation paths.
Progress is still being made on evolutionary models, but much remains
to be done, particularly with respect to the formation of single sdBs.
Asteroseismology is in an ideal position to test the different
formation scenarios, but the models of the interior must be made
precise enough so that they can reliably identify features of the
stellar interiors that remain distinct tracers of their evolutionary
history, even when their surface atmospheric parameters remain
indistinguishable \citep{hu08,hu09}.

But important questions about the interiors of the EHB stars still
remain. The current models predict that all EHB stars should pulsate
as soon as radiative levitation and gravitational settling have
accumulated sufficient iron group elements in the driving region.
But only 10\% of stars in the V361\,Hya instability region are found
to pulsate. While \citet{jeffery07} have speculated that this
accumulation can be disrupted by the vertical motion of strong
$p$-mode pulsations, it is hard to understand how this mechanism
can be reconciled with the time scales of the amplitude
variations that have been observed in these stars \citep{kilkenny10a}.
Another possibility was recently put forward by \citet{theado09},
who explore how iron-group enhanced layers lying on top light elements can
lead to convection through ``iron fingers''
(similar to the so-called ``salt fingers'' that are responsible for
thermohaline convection in the oceans).
Such convection can occur on timescales of a few thousand years,
much shorter than classical diffusion.
Asteroseismological models may have to include these types of diffusion 
in order to adequately represent the exited modes and amplitude
variation observed in EHB stars.

For V361\,Hya pulsators mode identifications have been made on a number
of stars with several different techniques that complement each other,
but not always producing identical answers. The high observed
mode density is a problem for frequency matching techniques,
especially when the predictive precision
of the models are only on the order of a few percent.
Mode identification from spectroscopy and multi-channel photometry has
the advantage that they can identify the $\ell$ order of modes without
any assumptions on the internal structure of the star. When combined
with radial velocity measurements this technique is particularly
powerful. More instruments that can support such observations will
be required in order to facilitate more extended application of
these techniques.
Three channel cameras can go a long way towards direct mode identification
on their own, but simultaneous radial velocity measurements would be
ideal. In principle, a low-resolution spectrograph can do this, but
current instruments are not well suited for photometric precision.
A minimum requirement is simultaneous observation of a reference star
in order to properly calibrate the photometry. This requires a derotator
in combination with an atmospheric dispersion corrector in the light
beam in order to avoid chromatic airmass effects. Such instruments
are currently not available for the sparsely populated fields where
we find the hot subdwarfs.

The low amplitudes and long periods make it very difficult to establish
detailed pulsation spectra for V1093\,Her stars,
and even when it can be done the high mode density makes
it difficult to assign modes to the observed frequencies.
But $g$-modes are particularly interesting because they probe deep into
the stellar interior. This is a significant challenge for the future
due to the long time-base required to reliably determine the longer
pulsation periods in these stars. The recently launched
{\em Kepler mission} \citep{kepler, borucki10}
provides an excellent opportunity for asteroseismology of compact
pulsators.

\acknowledgements
The author acknowledges funding from the European
Research Council under the European Community's Seventh Framework Programme
(FP7/2007--2013)/ERC grant agreement N$^{\underline{\mathrm o}}$\,227224
({\sc prosperity}), and from the Research Council of K.U.Leuven grant
agreement GOA/2008/04.

\bibliographystyle{an}
\bibliography{sdbrefs}

\end{document}